\documentclass[aps,twocolumn,flushleft,amsmath,amssymb,nofootinbib,superscriptaddress]{revtex4}

\usepackage{graphicx}
\usepackage{dcolumn} 
\usepackage{bm}      

\newcommand{\bq}{\begin{equation}}
\newcommand{\eq}{\end{equation}}
\newcommand{\ba}{\begin{eqnarray}}
\newcommand{\ea}{\end{eqnarray}}

\begin{document}
\title{Cosmological Implications of 5-dimensional Brans-Dicke Theory}
\author{Li-e Qiang}
\affiliation{National Astronomical Observatories, Chinese
Academy of Sciences, Beijing 100012, China}
\author{Yan Gong}
\affiliation{National Astronomical Observatories, Chinese
Academy of Sciences, Beijing 100012, China}
\affiliation{Graduate School of Chinese Academy of Sciences, Beijing 100049,
China}
\author{Yongge Ma}
\affiliation{Department of Physics, Beijing Normal
University, Beijing 100875, China}
\author{Xuelei Chen}
\affiliation{National Astronomical Observatories, Chinese
Academy of Sciences, Beijing 100012, China}
\affiliation{Center for High Energy Physics, Peking University, Beijing 100871,
China}

\begin{abstract}
The five dimensional Brans-Dicke theory naturally provides
two scalar fields by the Killing reduction mechanism.
These two scalar fields could account for the accelerated
expansion of the universe. We test this model and constrain its parameter
by using the type Ia supernova (SN Ia) data. We find that the best fit
value of the 5-dimensional Brans-Dicke coupling contant is
$\omega = -1.9$. This result is also consistent with other observations
such as the baryon acoustic oscillation (BAO).
\end{abstract}

\maketitle

\section{Introduction}

The expansion of the Universe is shown to be accelerating by the
observations of type Ia supernovae (SN Ia) \cite{Perlmutter99,
Riess98}. This is usually attributed to the contribution of an
unknown component, dubbed dark energy, which has negative
pressure and makes up about three quarters of the total cosmic density
(for recent measurements, see e.g. Ref.~\cite{Komatsu08,Gong08}).
The simplest model for dark
energy is the cosmological constant (CC), which is consistent
with most of the observations today. However, there are two big
problems for CC, i.e. the well-known ``fine tuning problem'' and
the ``coincidence problem''
. As alternatives, and also
to solve these two problems, many dynamical dark energy models with
scalar field have been proposed, such as
quintessence \cite{Wetterich88,Ratra88,FJ97,CDS98,CLW98,
Carroll98,ZMS99,ZMS991}, phantom \cite{Caldwell02},
quintom\cite{FWZ05}, K-essence \cite{Chiba00, Armendariz00}, tachyon
\cite{Padmanabhan02, Bagla03,Abramo03,Agui04,Guo04,Copeland05} and
so on. Nevertheless, in most cases the fundamental physical origin of these
 scalar fields remain unknown, but just added by hand.

In Ref. \cite{Qiang05}, by generalizing the Brans-Dicke theory to
five dimensions and exploring its effect on the 4-dimensional world,
another interesting approach to explain the cosmic accelerated
expansion was proposed. Under the condition that the extra dimension
is compact and sufficiently small, a spacelike Killing vector field
$\xi^a$ arises naturally, in which case the 5D Brans-Dicke theory
can be reduced to a 4D theory, such that the 4-metric is
coupled with two scalar fields $\phi$ and $\lambda$. Notes that here
the scalar fields in four dimension stem naturally from a fundamental theory of
gravity.  Considering
the hypersurface-orthogonal property of $\xi^a$, the line element in five dimension can take the form
as
\begin{equation}
ds^2=g_{\mu\nu}dx^{\mu}dx^{\nu}+\lambda dx^5dx^5,\label{metric}
\end{equation}
thus the scalar
$\lambda$ also plays the role of a ``scale factor'' of
the extra dimension. It was shown in Ref. \cite{Qiang05} that
these two scalar fields originated from the
Killing reduction of the 5D Brans-Dicke theory may lead to the
accelerated expansion of the universe. More detailed analysis is
desirable to check if the theory can match the current
observational data such as the SN Ia and the baryon acoustic
oscillation.

In this paper, we compare the predictions of the cosmic
expansion rate of this theory with the current cosmological
observations, and constrain the 5D Brans-Dicke theory by
means of the SN Ia data and the baryon acoustic oscillation (BAO)
measurements. This work is based on the fact that the 4-dimensional
gravitational constant $G$ varies extremely slowly with time
\cite{Turyshev04} in the current epoch. Thus we assume that $G$ is a constant at
the ``low redshift'', so that the accretion of the white
dwarf will not be affected, hence the luminosity and light curve of the 
observed SN Ia (redshifts range from 0 to 2) are not affected 
by the slow variations of $G$, and the SN
Ia can still be used as the standard candle.
Furthermore, if we assume that $G$ is almost a constant
throughout the history of the Universe, we
could also use the information from the large scale structure (e.g. BAO) to
perform the constraints. However, we should note that
$G=(\phi\lambda^{1/2}L)^{-1}$ \cite{Qiang05}, where $L$ is the
coordinate scale of the extra dimension. Although $G$
is almost constant in four dimension,
$G^{(5)}\sim\phi^{-1}$ is not necessarily a constant and can still evolve with time.

\section{theory}
The action of the five dimensional Brans-Dicke theory is given by
\ba
S_5&=&\int d^5x\sqrt{-g}(\phi
R^{(5)}-\frac{\omega}{\phi}g^{ab}(\nabla_a\phi)\nabla_b\phi)\nonumber\\&&+16\pi\int
d^5x\sqrt{-g}L_m^{(5)},
\ea
where $R^{(5)}$ is the curvature scalar
of the 5D metric $g_{ab}$, $\phi$
is the scalar field, $\omega$ is the coupling constant, and $L^{(5)}_m$
represents the Lagrangian of 5D matter fields.
Variation of this action gives the field equations
\ba
R_{ab}^{(5)}-\frac{1}{2}g_{ab}R^{(5)}&=&\frac{\omega}{\phi^{2}}
((\triangledown_a\phi)\triangledown_b\phi-\frac{1}{2}g_{ab}(\triangledown^c\phi)\triangledown_c\phi)
\nonumber\\
&&+\phi^{-1}(\triangledown_a\triangledown_b\phi-g_{ab}\triangledown^c\triangledown_c\phi)\nonumber\\
&&+8\pi\phi^{-1}T_{ab}^{(5)},\\
\triangledown^a\triangledown_a\phi&=&8\pi\frac{T^{(5)}}{4+3\omega},
\ea
where $T^{(5)}_{ab}$ represents the 5D energy
momentum tensor of matter fields and
$T^{(5)}=T^{(5)}_{ab}g^{ab}$.

The topology of the spacetime manifold is assumed to be $\mathbb{R}^4\times \mathbb{S}^1$,
and the extra dimension is confined into extremely small scales \cite{Hoyle}.
 Thus a Killing vector field $\xi^a$ arises naturally in
the low energy regime \cite{Blagojevic02}. Considering the case where
 $\xi^{a}$ is everywhere spacelike
and hypersurface orthogonal, the line element can be written down as
Eq.(\ref{metric}). The 4D Ricci tensor $R^{(4)}_{ab}$ of the
4-metric $h_{ab}$ and the scalar field $\lambda$ are related to the
5D Ricci tensor $R^{(5)}_{ab}$ by
\ba
R_{ab}^{(4)}=\frac{1}{2}\lambda^{-1}D_aD_b\lambda
-\frac{1}{4}\lambda^{-2}(D_a\lambda)D_b\lambda+h_a^ch_b^dR_{cd}^{(5)}
\ea
and
\ba D^{2}\lambda=\frac{1}{2}\lambda^{-1}(D^a\lambda)D_a\lambda
-2R_{ab}^{(5)}\xi^a\xi^b.
\ea
where $D_a$ is
the covariant derivative operator on the 4D spacetime obtained by
Killing reduction
$$D_eT^{b\ldots d}_{a\ldots c}=h^p_eh^m_a\ldots h^n_ch^b_r\ldots
h^d_s\nabla_pT^{r\ldots s}_{m\ldots n},$$ and  $D^2\equiv D^aD_a$.  After the Killing
reduction we obtain the 4-dimensional field equations
\cite{Qiang05}: \ba
G_{ab}^{(4)}&=&8\pi\phi^{-1}L^{-1}\lambda^{-\frac{1}{2}}T_{ab}^{(4)}
+\frac{1}{2}\lambda^{-1}(D_aD_b\lambda-h_{ab}D^cD_c\lambda)\nonumber\\
&&-\frac{1}{4}\lambda^{-2}((D_a\lambda)
D_b\lambda-h_{ab}(D^c\lambda) D_c\lambda)
  \nonumber\\
&&+\frac{\omega}{\phi^2}((D_a\phi)
  D_b\phi-\frac{1}{2}h_{ab}(D^c\phi)
D_c\phi)\nonumber\\
&&+\phi^{-1}(D_aD_b\phi-h_{ab}D^cD_c\phi)
\nonumber\\
&&-\frac{\phi^{-1}}{2}\lambda^{-1}h_{ab}(D^c\lambda)
D_c\phi,\label{G4}
\ea
\ba
D^aD_a\lambda&=&\frac{1}{2}\lambda^{-1}(D^a\lambda)
D_a\lambda-\phi^{-1}(D^a\lambda )D_a\phi\nonumber\\
&+&\frac{8\pi}{
L \lambda^{\frac{1}{2}}\phi}\left(\frac{2\omega+2}{4+3\omega}T^{(4)}
-\frac{4\omega+6}{4+3\omega}P\right),
\ea
and
\ba
D^aD_a\phi=-\frac{1}{2}\lambda^{-1}(D^c\lambda) D_c\phi+\frac{8\pi}{
L\lambda^{\frac{1}{2}}}\left(\frac{T^{(4)}+P}{4+3\omega}\right),\label{phi4}
\ea
which are equivalent with the 5D Brans-Dicke theory with the Killing symmetry.

In the homogeneous and isotropic universe described by
the 4D Robertson-Walker metric, Eqs.~(\ref{G4})-(\ref{phi4}) are
simplified as
\ba
\dot{H} &=& 2Hu+Hv+\frac{1}{2}uv-\frac{\omega}{2}u^2-8\pi G\frac{2\omega+3}{3\omega+4}\rho_m \label{H}\\
\dot{u} &=& -3Hu-u^2-\frac{1}{2}uv+8\pi G\frac{1}{3\omega+4}\rho_m \label{u} \\
\dot{v} &=& -3Hv-\frac{1}{2}v^2-uv+8\pi
G\frac{2\omega+2}{3\omega+4}\rho_m \label{v}\ea where
$H\equiv\dot{a}/a$ is the Hubble parameter,
$u\equiv\dot{\phi}/\phi$, $v\equiv\dot{\lambda}/\lambda$,
$\rho_m=\rho_{m_0}(1+z)^3$ is the matter density, and
$G=(\phi\lambda^{1/2}L)^{-1}$.

In the dynamical compactification model of Kaluza-Klein cosmology,
the extra dimensions contract while our 4-spacetime expands
\cite{Freund82,Mohammedi02,Darabi03}. We adopt this idea and assume
that the present Universe satisfies
$$a^3(t_0)\lambda^{n/2}(t_0)={\rm constant},$$
where $n$ is a positive real number, then we have \cite{Qiang05} \ba
v(t_0)=-\frac{6}{n}H_0, \qquad u(t_0)=\frac{3}{n}H_0. \ea
Substituting $\frac{d}{dt}=-H(1+z)\frac{d}{dz}$ into
Eq.(\ref{H})--Eq.(\ref{v}) with the present values of $H_0$, $u_0$ and
$v_0$, the evolutions of $H$, $u$ and $v$ according to the redshift could be solved
numerically.

\section{Observational Test}
We assume that the Universe is
flat, then the luminosity distance at a
redshift $z$ is given by
\bq
d_L(z)=(1+z)\int_0^z\frac{cdz'}{H(z')}.
\eq
The distance modulus is related to the luminosity distance by
\bq
\mu(z)=5\log_{10}d_L(z)+25.
\eq
In supernovae observation, the $\chi^2$ statistic is given by
\bq
\chi^2_{SN}=\sum_{i=1}^N\frac{(\mu_{obs}(z_i)-\mu_{th}(z_i))^2}{\sigma_i^2},
\eq
where $\mu_{th}(z_i), \mu_{obs}(z_i)$ are the
theoretically predicted and observed value
of the distance modulus at redshift $z_i$ respectively,
and $\sigma_i$ is the measurement error.

We use the SN Ia data recently published by the
Supernova Cosmology Project (SCP) team \cite{SCP08}.
This data set contains 307 selected SNe Ia, which includes
several widely used SNe Ia data set, such as the Hubble Space Telescope (HST)
\cite{Riess04, Riess06}, ``SuperNova Legacy Survey'' (SNLS) \cite{Astier05}
and the ``Equation of State: SupErNovae trace Cosmic Expansion''
(ESSENCE) \cite{ESSENCE07}.
Using the same analysis procedure and improved selection approach,
all of the sub-sets of data are analyzed to get a consistent
and high-quality ``Union'' data set, which
gives tighter and more reliable constraints.

The Markov Chain Monte Carlo (MCMC) technique is adopted to perform
the constraints. We generate eight MCMC chains and each chain
contains about two hundreds thousands simulated points after the
convergence has been reached. After the thinning process, there are
about 12000 points left to plot the marginalized probability
distribution function (PDF) and contour maps for the parameters in
our model. More details about our MCMC can be found in our
earlier paper \cite{Gong07}.

\section{Results}

\begin{figure}[htbp]
\includegraphics[scale = 0.35]{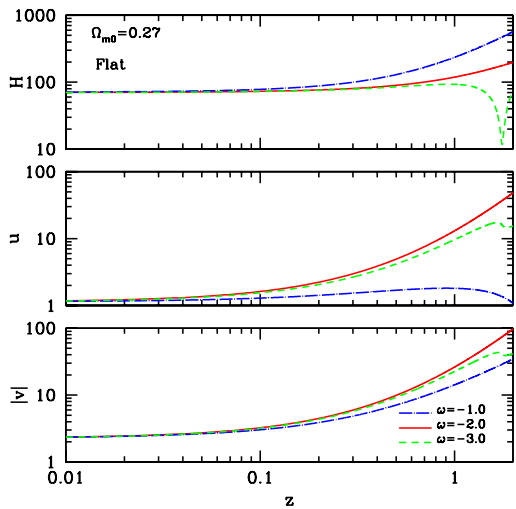}
\caption{\label{fig:H_u_v} The redshift evolution of $H$,
$u$ and $v$. We assume that the geometry of the Universe
is flat and fix $\Omega_{m_0}=0.27$, then plot the curves
for $\omega=-1$, $\omega=-2$ and $\omega=-3$ respectively.}
\end{figure}

\begin{figure}[hbtp]
\includegraphics[scale = 0.35]{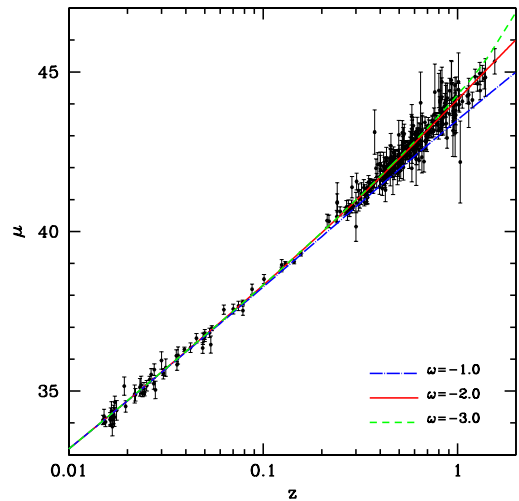}
\caption{\label{fig:mu}  Comparison of the distance moduli
between the models with $\omega=-1,-2,-3$
and the SCP SN Ia data set.}
\end{figure}

In Fig.\ref{fig:H_u_v}, we show the redshift 
evolution of $H$,$u$ and $|v|$ (where
$|v|=-v$). We set the present matter density parameter
$\Omega_{m_0}=0.27$. According to current solar system experiments,
the coupling constant $\omega$ of higher-dimensional Brans-Dicke
theories are constrained as $\omega \approx -(d-2)/(d-3)$ \cite{Klimek}, 
where $d$ is the
spatial dimension. For our case $d=2$, so $\omega \approx -2$. In the figure
we plotted the cases of $\omega=-1,-2,-3$.
As can be seen from the bottom and top panels of Fig.\ref{fig:H_u_v}, 
for different $\omega$
the extra dimension ``Hubble constant'' $\frac{v}{2}$ (note that
$v\equiv\dot{\lambda}/\lambda$) becomes more and more negative while
$H$ becomes more and more positive as the redshift goes up. This
indicates that the extra dimension is shrinking indeed while the
four visible dimensions are expanding. The cosmological implications
of this model can be seen more directly in Fig.~\ref{fig:mu}, where the 
distance moduli predicted by the theoretical model 
and the observed SN Ia data from SCP team
\cite{SCP08} are compared. Apparently the model prediction is 
in good agreement with data when $\omega=-2$. For $\omega=-1$ the
model acts as a matter-dominated universe, while for $\omega=-3$ as
a dark energy-dominated universe \cite{Copeland06}.

\begin{figure}[htbp]
\includegraphics[scale = 0.35]{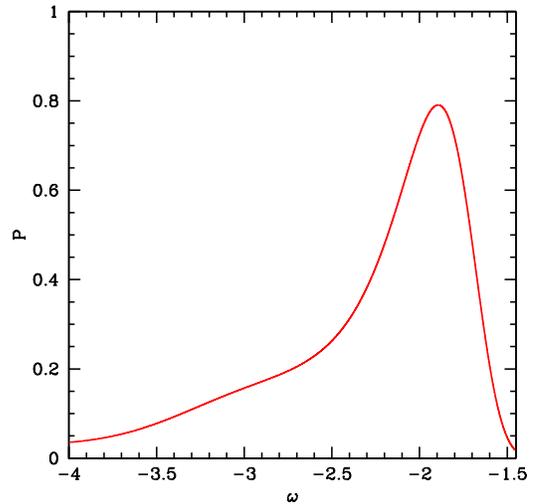}
\caption{\label{fig:w} The probability distribution function of $\omega$.
We find the best fit value for $\omega$ is about -1.9.
}
\end{figure}

We now investigate the constraint on the model.
The marginalized PDF of $\omega$ is shown in Fig.\ref{fig:w}. The best fit value
of $\omega$ is about -1.9. Note that $\omega \approx -2$ is required by 
solar system experiment \cite{Klimek}, and now we find that the 
best fit obtained with cosmological data  happens to give the 
same best fit $\omega$ value! This shows that our model 
predicts dark energy model naturally. 
The PDF decreases steeply when
$\omega>-1.9$ and gently when $\omega<-1.9$, so there
is also some probability for $\omega$ to get more negative values.

\begin{figure}[htbp]
\includegraphics[scale = 0.35]{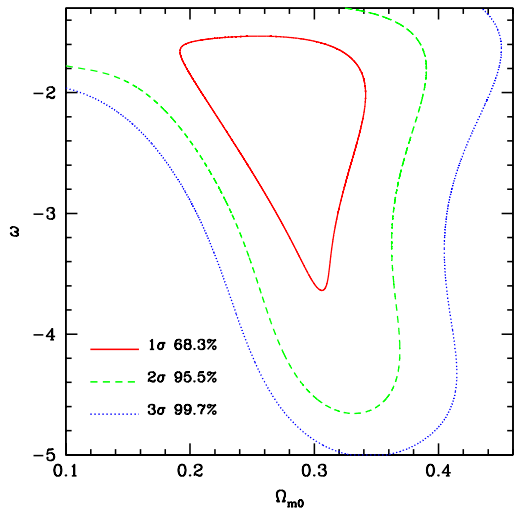}
\caption{\label{fig:m0_w} The contour map of $\Omega_{m_0}$ and $\omega$.
The joint best fit value of $\Omega_{m_0}$ and $\omega$ is around (0.27, -1.9).
The $1\sigma (68.3\%), 2\sigma (95.5\%)$ and $3\sigma (99.7\%)$ C.L. curves
are marked by red solid, green dashed and blue dotted lines
respectively.}
\end{figure}

In Fig.\ref{fig:m0_w}, we plot the contour map for $\Omega_{m_0}$ and $\omega$.
We find that the best fit value of $\Omega_{m_0}$ is around 0.27
which is consistent with other cosmological observations, e.g. 
cluster X-ray observations \cite{Allen08}. However, more negative values of
$\omega$ is also consistent with current observations. The
$95.5\%$ C.L.reaches -4.8, when $\Omega_{m_0}$ is
in the range of $0.24\sim 0.4$.

\section{Summary}

By considering a hypersurface-orthogonal spacelike Killing vector
field in the 5-dimensional spacetime, the 5D Brans-Dicke
theory can be reduced to a 4D theory with the 4-metric
coupled to two scalar fields. These two fields could naturally lead
to the accelerated expansion of the Universe.

We study the evolution of the two fields and compare the
expansion rate with SN Ia observations.  The two scalar
field would make the Universe evolve as if ``matter-dominate'' or ``dark
energy-dominate'' when $\omega$ is greater or less than -2.
We find that the model is in best agreement with the supernovae data
when the 5-dimensional coupling constant $\omega =-1.9 \approx -2$, 
which happens to be also the value required to satisfy the
solar system experiments. Furthermore, for this best fit value, 
the best fit $\Omega_{m_0}$ value is about 0.27, in good agreement with 
other independent measurements such as those derived from 
X-ray cluster observations. This work is based on the assumption
that the 4D gravitational constant $G$ varies extremely slowly
so that it can be regarded as a constant at "low redshift"
where the SN Ia data are
available. If we further assume $G$ does not change during the
whole history of the Universe, then other cosmological observations such as
BAO can also be used, we find that in this case the results are almost the same.

In conclusion, the 5-dimensional Brans-Dicke theory could
naturally provide two scalar fields which may cause the accelerated
expansion, the result is consistent with the SN Ia observation, hence
it is a candidate to explain the accelerated expansion of the
Universe.

\acknowledgments

Our MCMC chain computation was performed on the Supercomputing
Center of the Chinese Academy of Sciences and the Shanghai
Supercomputing Center. This work is supported by the National
Science Foundation of China under the Distinguished Young Scholar
Grant 10525314, the Key Project Grant 10533010, grant 10675019, by the Chinese
Academy of Sciences under grant KJCX3-SYW-N2, by the Ministry of
Science and Technology National Basic Science program (project 973)
under grant No.2007CB815401, by the Young Researcher Grant of 
National Astronomical Observatories, Chinese Academy of Sciences.


\newcommand\AL[3]{~Astron. Lett.{\bf ~#1}, #2~ (#3)}
\newcommand\AP[3]{~Astropart. Phys.{\bf ~#1}, #2~ (#3)}
\newcommand\AJ[3]{~Astron. J.{\bf ~#1}, #2~(#3)}
\newcommand\APJ[3]{~Astrophys. J.{\bf ~#1}, #2~ (#3)}
\newcommand\APJL[3]{~Astrophys. J. Lett. {\bf ~#1}, L#2~(#3)}
\newcommand\APJS[3]{~Astrophys. J. Suppl. Ser.{\bf ~#1}, #2~(#3)}
\newcommand\JCAP[3]{~JCAP. {\bf ~#1}, #2~ (#3)}
\newcommand\LRR[3]{~Living Rev. Relativity. {\bf ~#1}, #2~ (#3)}
\newcommand\MNRAS[3]{~Mon. Not. R. Astron. Soc.{\bf ~#1}, #2~(#3)}
\newcommand\MNRASL[3]{~Mon. Not. R. Astron. Soc.{\bf ~#1}, L#2~(#3)}
\newcommand\NPB[3]{~Nucl. Phys. B{\bf ~#1}, #2~(#3)}
\newcommand\PLB[3]{~Phys. Lett. B{\bf ~#1}, #2~(#3)}
\newcommand\PRL[3]{~Phys. Rev. Lett.{\bf ~#1}, #2~(#3)}
\newcommand\PR[3]{~Phys. Rep.{\bf ~#1}, #2~(#3)}
\newcommand\PRD[3]{~Phys. Rev. D{\bf ~#1}, #2~(#3)}
\newcommand\SJNP[3]{~Sov. J. Nucl. Phys.{\bf ~#1}, #2~(#3)}
\newcommand\ZPC[3]{~Z. Phys. C{\bf ~#1}, #2~(#3)}

\end{document}